\documentclass[aps,prl,reprint,superscriptaddress]{revtex4-1}
\usepackage{graphicx}
\usepackage{color}

\usepackage[utf8]{inputenc}

\usepackage{epstopdf}
\begin{document}

\title{Simultaneous conduction and valence band quantisation in ultra-shallow, high density doping profiles in semiconductors}

\author{F.~Mazzola}
\affiliation{Center for Quantum Spintronics, Department of Physics, Norwegian University of Science and Technology, NO-7491 Trondheim, Norway }
\author{J.~W.~Wells}
\email{quantum.wells@gmail.com}
\affiliation{Center for Quantum Spintronics, Department of Physics, Norwegian University of Science and Technology, NO-7491 Trondheim, Norway }
\author{A.~C.~Pakpour-Tabrizi}
\affiliation{London Centre for Nanotechnology and Department of Electronic and Electrical Engineering, University College London, 17-19 Gordon Street, London WC1H 0AH, U.K.}
\author{R.~B.~Jackman}
\affiliation{London Centre for Nanotechnology and Department of Electronic and Electrical Engineering, University College London, 17-19 Gordon Street, London WC1H 0AH, U.K.}
\author{B.~Thiagarajan}
\affiliation{MAX-lab, PO Box 118, S-22100 Lund, Sweden}
\author{Ph.~Hofmann}
\affiliation{ Department of Physics and Astronomy and Interdisciplinary Nanoscience Center (iNANO), Aarhus University Ny Munkegade 120 DK-8000 Aarhus}
\author{J.~A.~Miwa}
\affiliation{ Department of Physics and Astronomy and Interdisciplinary Nanoscience Center (iNANO), Aarhus University Ny Munkegade 120 DK-8000 Aarhus}

\date{\today}

\begin{abstract}

We demonstrate simultaneous quantisation of conduction band (CB) and valence band (VB) states in silicon using ultra-shallow, high density, phosphorus doping profiles (so-called Si:P $\delta$-layers). We show that, in addition to the well known quantisation of CB states within the dopant plane, the confinement of VB-derived states between the sub-surface P dopant layer and the Si surface gives rise to a simultaneous quantisation of VB states in this narrow region.  We also show that the VB quantisation can be explained using a simple particle-in-a-box model, and that the number and energy separation of the quantised VB states depend on the depth of the P dopant layer beneath the Si surface. Since the quantised CB states do not show a strong dependence on the dopant depth (but rather on the dopant density), it is straightforward to exhibit control over the properties of the quantised CB and VB states independently of each other by choosing the dopant density and depth accordingly, thus offering new possibilities for engineering quantum matter. 

\end{abstract}

 \maketitle
 
There has been a surge of interest in two-dimensional (2D) materials due to their remarkable quantum properties. Graphene and layered transition metal dichalcogenides are just two examples of such materials that have been recently studied and proffered as advantageous for developing quantum electronic devices \cite{Novoselov:2004,Novoselov:2007, Mak:2014, Xu:2014}. A rather unique branch of the 2D material family are ultra-shallow, high-density, doping profiles in semiconductors, so-called $\delta$-layers.  In particular, phosphorus $\delta$-layers in silicon (Si:P $\delta$-layers)  combined with atomically precise lithography have led to recent technological successes towards scalable qubit architectures \cite{Fuechsle:2012,Weber:2012, Hill:2015, McKibbin:2013}. It has been demonstrated that P donors, which can act as qubits, in Si have long spin lifetimes \cite{Suzuki:2011,Wolfowicz:2013} which are essential for spin-based quantum calculations. Importantly, Si:P $\delta$-layers can be readily synthesized -- they are comprised of a  Si(001) substrate with a high density P dopant profile situated a few nanometers beneath an epitaxial grown Si encapsulation layer -- and they are potentially straightforward to integrate into existing Si-based technology.  Both the dopant layer and the encapsulation layer can be easily modified during the growth process \cite{Goh:2004, Goh:2006, Goh:2009}, and it is this flexibility that  makes $\delta$-layers so promising, not only for enhancing the performance of quantum electronic devices, but for engineering new 2D materials with new capabilities.

The confinement of a high density, atomically thin layer of P atoms beneath the surface abruptly changes the potential within the Si crystal.  This brings about strong bending of the conduction band (CB) and valence band (VB) around the dopant plane, leading to strong confinement of the silicon CB. This strong confinement results in lowering and discretisation of the CB and consequently gives the system metallic character  \cite{Carter:2009a, Carter:2011,Miwa:2013, Polley:2012a,Polley:2013a}. These CB states have been studied in considerable detail  \cite{Miwa:2013, Miwa:2014a, Mazzola:2014, Mazzola:2014a}, and it has been determined that their binding energy and energy separation (so-called valley splitting) can be effectively controlled and tuned by varying the P doping density and/or depth profile \cite{Mazzola:2016}.

It is not only essential for device operation and performance that the quantised CB states can be tuned and controlled but also their VB counterparts. We demonstrate a general method based on $\delta$-doping to realise simultaneous quantisation of CB and VB electrons by structuring the band bending at the nanoscale. We show, using angle-resolved photoemission spectroscopy (ARPES), that quantised VB states  arise from confinement between the P dopant layer and surface of the Si encapsulation.  We verify that these quantised VB states can be tuned by varying the thickness of the Si encapsulation.  This capability promises new prospects in engineering quantum matter, for example, the possibility of controlling carrier lifetimes by modifying the interaction between quantised CB and VB states.

ARPES measurements were performed at the I4 beamline at the MAX-III synchrotron radiation source \cite{Jensen:1997}.  The energy and momentum resolutions were better than $40$~meV and $0.02$~\AA$^{-1}$, respectively. The base pressure in the analysis chamber was $\approx 5 \times 10^{-10}$~mbar, and the temperature of the sample was maintained at room temperature throughout data acquisition.  The Si:P $\delta$-layer samples were prepared by growing epitaxial Si (thicknesses from $1$ to $4$~nm) on top of  $\approx$0.25 monolayers of P atoms incorporated in the topmost layer of a clean Si(001) substrate; a detailed recipe can be found in Ref.\ \onlinecite{Miwa:2013}. The arrangement of incorporated P atoms in the Si substrate has been investigated by combined atom-resolved scanning tunnelling microscopy \cite{Curson:2004} and density functional theory \cite{Wilson:2006}. The results of these studies suggest the incorporated P atoms exhibit some short- but no  long-range ordering. Control samples were measured for comparison, and fabricated by growing a similar amount of epitaxial Si directly on the clean Si(001) substrate \textit{without} the inclusion of a P-rich layer.

\begin{figure}
\centering
\includegraphics [width=\columnwidth]{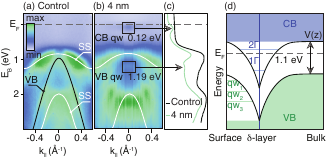}
\caption{Simultaneous quantisation for CB and VB states in silicon. (a) ARPES data for the control sample, (b) corresponding ARPES data of a Si:P $\delta$-layer sample with a 4\,nm encapsulation thickness, CB and VB states indicated. (c) Momentum integrated EDCs to emphasise the differences between the two samples. (d) Band bending schematic of a Si:P $\delta$-layer. The resulting potential, $V(z)$, is shown by the black line and confines the CB band electrons to give rise to the states labelled 1$\Gamma$ and 2$\Gamma$. Recovery of the potential well to the surface leads to quantised states confined to the Si encapsulation layer: qw$_1$, qw$_2$ and qw$_3$. The blue and green shaded areas represent the continuum of CB and VB bulk states where these quantised states \textit{cannot} form.} \label{schem1}
\end{figure}

The ARPES acquisitions of a control sample and a Si:P $\delta$-layer sample with an approximate 4 nm thick Si encapsulation layer are presented in Fig.\ 1(a) and (b), respectively. The two samples have similar spectral features: the VB dispersion around the $\bar{\mathrm{\Gamma}}$ point of the 2D Brillouin zone and surface states (SS) are consistent with bulk and surface states previously reported for electronic structure measurements of Si(001) with a 2$\times$1 reconstructed surface \cite{Johansson:1990a, Johansson:1990}. 

The Fermi level ($E_F$) lies within the 1.1\,eV band gap and situated in close proximity to the conduction band minimum (CBM); confirming the \textit{n}-type doping. There are notable differences between the two samples: an additional feature near $E_F$ and extra bands in the VB region --- marked by the black rectangles --- can be seen in the ARPES data of the Si:P $\delta$-layer sample shown in Fig.\ \ref{schem1}(b). These differences are prominent  in Fig.\ \ref{schem1}(c) where energy distribution curves (EDCs), integrated over a momentum range of -0.15 to 0.15~\AA$^{-1}$, are plotted.  The additional states appear as peaks at binding energies of 0.12\,eV and 1.19\,eV for the Si:P $\delta$-layer sample (green curve) and are noticeably absent in the control sample (black curve). 

We use the band bending diagram of a Si:P $\delta$-layer in Fig.\ \ref{schem1}(d) to illustrate the origin of these additional states.  As we go from bulk to surface, i.e. from right to left across the diagram, the bulk CB becomes partially occupied in the region around the high density P dopant plane thereby creating a confined metallic layer.  The CB states which are bound by the Coulomb-like potential well are labelled 1$\Gamma$ and 2$\Gamma$. Whilst these electronic states have already been studied in detail by ARPES \cite{Miwa:2014a}, an understanding of the nature and origin of the extra bands that are visible in the VB region is lacking.  Previous ARPES measurements have shown that both the CB and VB states are non-dispersing with photon energy, firmly establishing the 2D character of these co-existing states \cite{Miwa:2013}. ARPES acquisitions at photon energy of 36\,eV are only presented here, as the intensity of the CB states is known to be enhanced at this energy \cite{Miwa:2013, Mazzola:2014a}. 

If VB states should exist between the surface and the $\delta$-layer, they too must be strongly confined since both the surface and the $\delta$-layer act as a barrier (see left side of Fig.\ \ref{schem1}(d)). Therefore, the hole-like bands of the VB become trapped like a particle-in-a-box, where the confinement width is dictated by the depth of the $\delta$-layer beneath the surface and the confinement potential is dictated by the Fermi level pinning at the surface and at the $\delta$-layer. All of these parameters can be controlled during the sample growth. 

\begin{figure*}[ht]
\centering
\includegraphics [width=2\columnwidth]{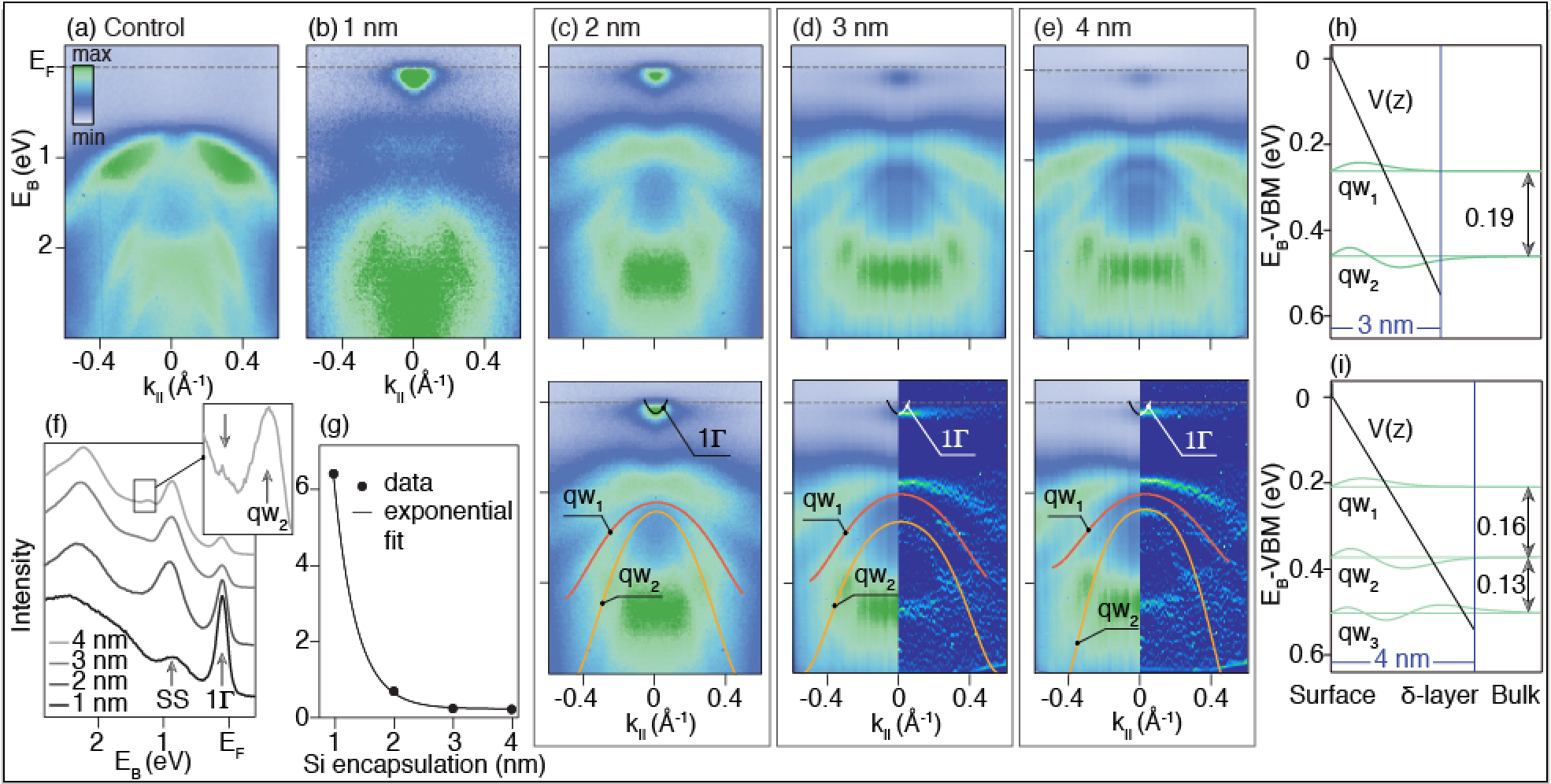}
\caption{Evolution of the quantised VB states with encapsulation thickness. (a) ARPES data for the control Si sample. (b -- e) Si:P $\delta$-layer ARPES spectra acquired for different Si encapsulation thicknesses ranging from 1\,nm to 4\,nm. The CB state at the Fermi level (1$\Gamma$) becomes gradually weaker with increasing Si encapsulation thickness while the states within the VB region become more intense. For panels (c -- e), the ARPES spectra are shown twice with salient features marked and labelled on the spectra displayed in the lower panels. To enhance the visibility of the quantised VB states, the curvature method   \cite{Zhang:2011}  was applied and the results  are presented on right sides of the lower panels of (d) and (e).  The spectra for the 1\,nm -- 4\,nm encapsulation thicknesses have been left-right symmetrised, while the control sample has not. An even sixth-order polynomial was used to fit the quantised VB states, qw$_1$ (orange) and qw$_2$ (yellow), for the data shown in panels (c -- e) \cite{SM1}.  (f) Momentum integrated EDCs for Si:P $\delta$-layers with different encapsulation thicknesses. The positions of the CB (1$\Gamma$) and SS are indicated. The inset shows an enlarged region of the 4\,nm thick Si encapsulation data where the qw$_2$ state is readily visible. Adjacent to the qw$_2$ state, an unlabelled arrow marks the location of small peak which may be due to a third quantised valence band state. (g) The intensity ratio of CB to VB states for each of the Si:P $\delta$-layer samples. (h) and (i) Numerically obtained solutions to the Schr\"{o}dinger equation for a linear potential well ($V(z)$, black line) for 3\,nm and 4\,nm Si encapsulation layers, respectively. The calculated eigenstates are marked by the green curves and the separation energies (in eV) marked by the double-headed arrows.} \label{schem2}
\end{figure*}

We have explored the influence of Si encapsulation layer thickness on the quantisation of CB and VB states using ARPES. First of all, we want to study the thickness dependent photoemission intensity, as this will give information about whether a certain electronic state is bulk or surface derived.  In Fig.\ \ref{schem2}(b-e) we consider the following Si encapsulation thicknesses: 1\,nm, 2\,nm, 3\,nm, 4\,nm and compare them with the ARPES measurements for the control sample shown in Fig.\ \ref{schem2}(a).  At first glance, all of the ARPES spectra of the different Si:P $\delta$-layer samples appear qualitatively similar to each other. A pronounced difference is the diminishing spectral weight of 1$\Gamma$ near the $E_F$ for Si:P $\delta$-layers with thicker encapsulation. Since 1$\Gamma$ originates from the P dopant layer situated beneath the surface, it is expected that the signal intensity gets weaker for thicker Si encapsulations.  In Fig.\ \ref{schem2}(f), EDCs (integrated over a momentum range of -0.15 to 0.15\,\AA$^{-1}$) are plotted for the control sample and the four Si:P $\delta$-layers.  In this manner, the peak intensities of both the CB and VB quantised states, marked by the arrows, can be directly compared. For increasing Si encapsulation thicknesses, a decrease in the intensity of the quantised CB peak corresponds to an increase in the intensity of the quantised VB states.  This is confirmed in Fig.\ \ref{schem2}(g) where the spectral intensity of the CB is plotted (relative to the intensity of the VB states, i.e. $I_\delta$/$I_{VB}$), as a function of Si encapsulation thickness and shows an exponential suppression for photoelectrons emitted from deeper P dopant layers. Whilst the sub-surface origin of the quantised CB states was known, this analysis suggests that the quantised VB states exist up to the surface.

By increasing the thickness of the encapsulation from 3\,nm to 4\,nm, the energy separation between the quantised VB states decreases; compare qw$_1$ (orange) and  qw$_2$ (yellow) in Fig.\ \ref{schem2}(d and e). This trend can be explained by a particle-in-a-box picture: as the width of the box, or in this case the thickness of the Si encapsulation, is increased, the energies of the quantum states are lowered. We note that the energy separation between the quantised VB states for the 2\,nm encapsulation thickness (Fig.\ \ref{schem2}(c)) does not follow this trend. This exception may be due to the complex interaction of the SS, located at  $E_b \approx$ 1\,eV, with the quantised VB states. While the physical extent of the SS wave function is relatively shallow \cite{Yengui:2015aa}, a broadening and shifting could still be expected for a sufficiently small spatial separation of the SS and the quantised VB states. 

The dispersion of the quantised VB states was fitted using an even sixth-order polynomial (orange and yellow curves in Fig.\ \ref{schem2}), and their effective masses and uncertainties estimated \cite{SM2}.  We expect the different encapsulation thicknesses to have a small affect on the effective masses of the quantised VB states. Given the associated uncertainties, the effective masses for the qw$_1$ state are in agreement with the heavy-hole state in the bulk VB of Si. The effective masses for the qw$_2$ state are less, but also probably derived from the bulk heavy-holes since their effective masses are more similar to that of the bulk heavy-hole state than the bulk light-hole state.

Additional confirmation that the extra features in the VB region are quantised VB states confined within the Si encapsulation layer is provided by our numerical model for solving the Schr\"{o}dinger equation presented in Fig.\ \ref{schem2}(h and i). For simplicity we only consider a linear potential ($V(z)$, black line) between the dopant layer and the surface. The approximation is crude but reproduces the quantised VB states seen in the ARPES measurements of Fig.\ \ref{schem2}. It is worth noting that we only apply our model to Si:P $\delta$-layers with the thickest encapsulation layers studied here, i.e. $3$~nm and $4$~nm, since the quantised VB states and the surface states are well separated for these cases thereby facilitating the comparison between data and model. The interaction of the SS with the quantised VB states, for the $1$~nm and $2$~nm cases, is simply not captured in this  model that assumes a quantum well with the same boundary conditions for every thickness.

In our model by increasing the thickness from 3 to 4~nm, the number of solutions to the Schr\"{o}dinger equation increases from two to three.  For the 3\,nm case, the two calculated states  are assigned to the qw$_1$ and qw$_2$ states observed in the experiment; see Fig.\ \ref{schem2}(d). The experimental data in Fig.\ \ref{schem2}(e) shows a weak hint of a third qw$_3$ state expected for the 4\,nm case but the intensity of the signal is weak and comparable to the background. The reduced intensity of the state may also be a result of its wave function being less localised at the surface (compared to qw$_1$ \& qw$_2$) as illustrated in Fig.\ \ref{schem2}(i), or due to the fact that the photoionisation cross-section of this state is lower at this photon energy \cite{Miwa:2013} (or, most likely, both effects might play a role).  

We extracted EDCs, integrated over a finite momentum range, for Si:P $\delta$-layers with different encapsulation thicknesses to investigate further this possible qw$_3$ state.  In the inset of Fig.\ \ref{schem2}(f), the qw$_2$ state is readily visible for the 4\,nm thick Si encapsulation data, and adjacent to this state, there is a small peak where the qw$_3$ state may be expected.  

The energy separations between the quantised VB states are determined from the numerical model to be: qw$_2$-qw$_1$=0.19\,eV for the $3$~nm thick Si encapsulation, and qw$_2$-qw$_1$=0.16\,eV and qw$_3$-qw$_2$=0.13\,eV for the $4$~nm thick layer. From the experimental data we measure an energy separation between the two lowest lying states to be qw$_2$-qw$_1$=$0.30\pm0.17$~eV for the 3~nm thick Si encapsulation and qw$_2$-qw$_1$=$0.17\pm0.12$~eV for the 4~nm thick layer, respectively (3~nm: qw$_1=1.02\pm0.08$~eV and qw$_2=1.32\pm 0.09$~eV, $4$~nm: qw$_1=1.02\pm0.08$~eV and qw$_2=1.19\pm 0.04$~eV).  The experimental values for all the quantised VB states are different from the ones extracted numerically, however the general trend holds for the thicker encapsulation thicknesses: (i) a shift of the quantised states toward lower binding energy and (ii) a decrease in the energy separation between higher to lower lying states for increasing Si encapsulation thickness is observed. This supports the notion that the quantised VB states originate from confinement in the Si encapsulation layer. We expect that a more accurate model for the doping potential and its recovery to the surface, including the influence of the SS wave function, might be able to give more realistic energy separations. 

Simultaneous quantisation of CB and VB has not been demonstrated in common semiconductors, previously, a special case of simultaneous quantisation of the CB and VB has been reported for the topological insulator Bi$_2$Se$_3$ \cite{Bianchi:2011}.  The adsorption of CO gas on the Bi$_2$Se$_3$ surface induces a similar downward band bending of the CB and the formation of quantised CB states. However, the quanitsed VB has quite another origin: Bi$_2$Se$_3$ has a peculiar valence electronic structure near the centre of its surface Brillouin zone  --- in this region the upper VB only  exists in a narrow ($\approx$200~meV) energy window --- and thus downward bending of the VB can also lead to quantised states. The origin of the CB and VB quantisation is completely different in a $\delta$-layer since the simultaneous quantisation of the CB and VB is purely artificial: it is dictated by the type, density and profile of the dopant layer and unlike Bi$_2$Se$_3$ is not an innate and unusual property of the bulk material. Artificially induced quantisation of the CB and VB by $\delta$-doping offers the realisation of the same effect in a wide spectrum of semiconductor hosts. 

The properties of the band-bending in $\delta$-layers can be easily modified during the growth process, and as a result quantisation of the CB and VB can be controlled and tuned.   We can, for example, also occupy the 2$\Gamma$ state so that it is situated below the $E_{F}$ \cite{Miwa:2014a,Mazzola:2016}, by either increasing the P dopant density or broadening the P dopant profile of the $\delta$-layer.  The surface of the Si encapsulation layer will similarly impact the quantisation of the VB states as different surface terminations or surface adsorbates can alter  the $E_F$ pinning  at the surface, and thus modify the degree of band bending between the dopant layer and the surface. 

The situation of simultaneous quantisation of electron and hole states is a rather unusual effect \cite{Bianchi:2011}, never observed before in traditional doped semiconductors. This effect provides the appealing prospect of controlling the lifetime of carriers, creating additional channels to generate electron-hole pair recombinations in the CB, mediated by electronic transitions from the VB, potentially controlled by a biased top gate (analogous to a field effect transistor). That is, to mediate transitions between the CB and VB by tuning the potential landscape in which these states reside by modification of the dopant layer and the surface termination. For example, surface doping would directly influence the barrier potential responsible for the near-surface quantised VB states, and thus directly influence their energy, but would have a minimal influence on the sub-surface quantised CB (and bulk VB), for which the $\delta$-layer and bulk doping densities, respectively, determine the Fermi level pinning. Thus, by modifying the surface potential, it should be possible to deliberately align (or misalign) the energies of the quanitised VB and CB states so as to exhibit control of their interaction (and therefore lifetime). The flexibility that these $\delta$-layers offer could be expected to play a major role in the performance of quantum electronic devices. 

In summary, simultaneous quantisation of the CB and VB states of Si:P $\delta$-layers has been experimentally verified using ARPES. The origins of these quantised states are  different: the CB states arise from the potential well induced by the ultra-dense dopant layer whereas the VB states originate from confinement between the potential well created by the dopant layer, and the sample surface.  All of the relevant properties of both the dopant and encapsulation layers can be easily controlled and modified during the $\delta$-layer growth process; not only providing the ability to exhibit control the quantisation of CB and VB states, but also offering the intriguing possibility of influencing lifetimes within the  $\delta$-layer structure, thereby opening up new possibilities for engineering quantum materials with new capabilities.\\

\noindent\textbf{Acknowledgements:} We acknowledge Johan Adell for  support at the I4 beamline at MAX-III. Partial funding for this work was obtained through the Norwegian PhD Network on Nanotechnology for Microsystems sponsored by the Research Council of Norway, Division for Science under contract no. 221860/F40. J.A.M.\ acknowledges support from the  Danish Council for Independent Research, Natural Sciences under the \textit{Sapere Aude} program (Grant no.\ DFF-6108-00409) and the Aarhus University Research Foundation. This work was supported by VILLUM FONDEN via the Centre of Excellence for Dirac Materials (Grant No. 11744) and partly supported by the Research Council of Norway through its Centres of Excellence funding scheme, project number 262633, ``QuSpin'', and through the Fripro program, project number 250985 ``FunTopoMat''.


\begin{thebibliography}{32}
\expandafter\ifx\csname natexlab\endcsname\relax\def\natexlab#1{#1}\fi
\expandafter\ifx\csname bibnamefont\endcsname\relax
  \def\bibnamefont#1{#1}\fi
\expandafter\ifx\csname bibfnamefont\endcsname\relax
  \def\bibfnamefont#1{#1}\fi
\expandafter\ifx\csname citenamefont\endcsname\relax
  \def\citenamefont#1{#1}\fi
\expandafter\ifx\csname url\endcsname\relax
  \def\url#1{\texttt{#1}}\fi
\expandafter\ifx\csname urlprefix\endcsname\relax\def\urlprefix{URL }\fi
\providecommand{\bibinfo}[2]{#2}
\providecommand{\eprint}[2][]{\url{#2}}

\bibitem[{\citenamefont{Novoselov et~al.}(2004)\citenamefont{Novoselov, Geim,
  Morozov, Jiang, Zhang, Dubonos, Grigorieva, and Firsov}}]{Novoselov:2004}
\bibinfo{author}{\bibfnamefont{K.~S.} \bibnamefont{Novoselov}},
  \bibinfo{author}{\bibfnamefont{A.~K.} \bibnamefont{Geim}},
  \bibinfo{author}{\bibfnamefont{S.~V.} \bibnamefont{Morozov}},
  \bibinfo{author}{\bibfnamefont{D.}~\bibnamefont{Jiang}},
  \bibinfo{author}{\bibfnamefont{Y.}~\bibnamefont{Zhang}},
  \bibinfo{author}{\bibfnamefont{S.~V.} \bibnamefont{Dubonos}},
  \bibinfo{author}{\bibfnamefont{I.~V.} \bibnamefont{Grigorieva}},
  \bibnamefont{and} \bibinfo{author}{\bibfnamefont{A.~A.}
  \bibnamefont{Firsov}}, \bibinfo{journal}{Science}
  \textbf{\bibinfo{volume}{306}}, \bibinfo{pages}{666} (\bibinfo{year}{2004}).

\bibitem[{\citenamefont{Novoselov et~al.}(2007)\citenamefont{Novoselov, Jiang,
  Zhang, Morozov, Stormer, Zeitler, Maan, Boebinger, Kim, and
  Geim}}]{Novoselov:2007}
\bibinfo{author}{\bibfnamefont{K.~S.} \bibnamefont{Novoselov}},
  \bibinfo{author}{\bibfnamefont{Z.}~\bibnamefont{Jiang}},
  \bibinfo{author}{\bibfnamefont{Y.}~\bibnamefont{Zhang}},
  \bibinfo{author}{\bibfnamefont{S.~V.} \bibnamefont{Morozov}},
  \bibinfo{author}{\bibfnamefont{H.~L.} \bibnamefont{Stormer}},
  \bibinfo{author}{\bibfnamefont{U.}~\bibnamefont{Zeitler}},
  \bibinfo{author}{\bibfnamefont{J.~C.} \bibnamefont{Maan}},
  \bibinfo{author}{\bibfnamefont{G.~S.} \bibnamefont{Boebinger}},
  \bibinfo{author}{\bibfnamefont{P.}~\bibnamefont{Kim}}, \bibnamefont{and}
  \bibinfo{author}{\bibfnamefont{A.~K.} \bibnamefont{Geim}},
  \bibinfo{journal}{Science} \textbf{\bibinfo{volume}{315}},
  \bibinfo{pages}{1379} (\bibinfo{year}{2007}).

\bibitem[{\citenamefont{Mak et~al.}(2014)\citenamefont{Mak, McGill, Park, and
  McEuen}}]{Mak:2014}
\bibinfo{author}{\bibfnamefont{K.~F.} \bibnamefont{Mak}},
  \bibinfo{author}{\bibfnamefont{K.~L.} \bibnamefont{McGill}},
  \bibinfo{author}{\bibfnamefont{J.}~\bibnamefont{Park}}, \bibnamefont{and}
  \bibinfo{author}{\bibfnamefont{P.~L.} \bibnamefont{McEuen}},
  \bibinfo{journal}{Science} \textbf{\bibinfo{volume}{344}},
  \bibinfo{pages}{1489} (\bibinfo{year}{2014}).

\bibitem[{\citenamefont{Xu et~al.}(2014)\citenamefont{Xu, Yao, Xiao, and
  Heinz}}]{Xu:2014}
\bibinfo{author}{\bibfnamefont{X.}~\bibnamefont{Xu}},
  \bibinfo{author}{\bibfnamefont{W.}~\bibnamefont{Yao}},
  \bibinfo{author}{\bibfnamefont{D.}~\bibnamefont{Xiao}}, \bibnamefont{and}
  \bibinfo{author}{\bibfnamefont{T.~F.} \bibnamefont{Heinz}},
  \bibinfo{journal}{Nat. Phys.} \textbf{\bibinfo{volume}{10}},
  \bibinfo{pages}{343} (\bibinfo{year}{2014}).

\bibitem[{\citenamefont{Fuechsle et~al.}(2012)\citenamefont{Fuechsle, Miwa,
  Mahapatra, Ryu, Lee, Warschkow, Hollenberg, Klimeck, and
  Simmons}}]{Fuechsle:2012}
\bibinfo{author}{\bibfnamefont{M.}~\bibnamefont{Fuechsle}},
  \bibinfo{author}{\bibfnamefont{J.~A.} \bibnamefont{Miwa}},
  \bibinfo{author}{\bibfnamefont{S.}~\bibnamefont{Mahapatra}},
  \bibinfo{author}{\bibfnamefont{H.}~\bibnamefont{Ryu}},
  \bibinfo{author}{\bibfnamefont{S.}~\bibnamefont{Lee}},
  \bibinfo{author}{\bibfnamefont{O.}~\bibnamefont{Warschkow}},
  \bibinfo{author}{\bibfnamefont{L.~C.~L.} \bibnamefont{Hollenberg}},
  \bibinfo{author}{\bibfnamefont{G.}~\bibnamefont{Klimeck}}, \bibnamefont{and}
  \bibinfo{author}{\bibfnamefont{M.~Y.} \bibnamefont{Simmons}},
  \bibinfo{journal}{Nat. Nanotechnol.} \textbf{\bibinfo{volume}{7}},
  \bibinfo{pages}{242} (\bibinfo{year}{2012}).

\bibitem[{\citenamefont{Weber et~al.}(2012)\citenamefont{Weber, Mahapatra, Ryu,
  Lee, Fuhrer, Reusch, Thompson, Lee, Klimeck, Hollenberg et~al.}}]{Weber:2012}
\bibinfo{author}{\bibfnamefont{B.}~\bibnamefont{Weber}},
  \bibinfo{author}{\bibfnamefont{S.}~\bibnamefont{Mahapatra}},
  \bibinfo{author}{\bibfnamefont{H.}~\bibnamefont{Ryu}},
  \bibinfo{author}{\bibfnamefont{S.}~\bibnamefont{Lee}},
  \bibinfo{author}{\bibfnamefont{A.}~\bibnamefont{Fuhrer}},
  \bibinfo{author}{\bibfnamefont{T.~C.~G.} \bibnamefont{Reusch}},
  \bibinfo{author}{\bibfnamefont{D.~L.} \bibnamefont{Thompson}},
  \bibinfo{author}{\bibfnamefont{W.~C.~T.} \bibnamefont{Lee}},
  \bibinfo{author}{\bibfnamefont{G.}~\bibnamefont{Klimeck}},
  \bibinfo{author}{\bibfnamefont{L.~C.~L.} \bibnamefont{Hollenberg}},
  \bibnamefont{et~al.}, \bibinfo{journal}{Science}
  \textbf{\bibinfo{volume}{335}}, \bibinfo{pages}{64} (\bibinfo{year}{2012}).

\bibitem[{\citenamefont{Hill et~al.}(2015)\citenamefont{Hill, Peretz, Hile,
  House, Fuechsle, Rogge, Simmons, and Hollenberg}}]{Hill:2015}
\bibinfo{author}{\bibfnamefont{C.~D.} \bibnamefont{Hill}},
  \bibinfo{author}{\bibfnamefont{E.}~\bibnamefont{Peretz}},
  \bibinfo{author}{\bibfnamefont{S.~J.} \bibnamefont{Hile}},
  \bibinfo{author}{\bibfnamefont{M.~G.} \bibnamefont{House}},
  \bibinfo{author}{\bibfnamefont{M.}~\bibnamefont{Fuechsle}},
  \bibinfo{author}{\bibfnamefont{S.}~\bibnamefont{Rogge}},
  \bibinfo{author}{\bibfnamefont{M.~Y.} \bibnamefont{Simmons}},
  \bibnamefont{and} \bibinfo{author}{\bibfnamefont{L.~C.~L.}
  \bibnamefont{Hollenberg}}, \bibinfo{journal}{Sci. Adv.}
  \textbf{\bibinfo{volume}{1}}, \bibinfo{pages}{e1500707}
  (\bibinfo{year}{2015}).

\bibitem[{\citenamefont{McKibbin et~al.}(2013)\citenamefont{McKibbin,
  Scappucci, Pok, and Simmons}}]{McKibbin:2013}
\bibinfo{author}{\bibfnamefont{S.~R.} \bibnamefont{McKibbin}},
  \bibinfo{author}{\bibfnamefont{G.}~\bibnamefont{Scappucci}},
  \bibinfo{author}{\bibfnamefont{W.}~\bibnamefont{Pok}}, \bibnamefont{and}
  \bibinfo{author}{\bibfnamefont{M.~Y.} \bibnamefont{Simmons}},
  \bibinfo{journal}{Nanotechnology} \textbf{\bibinfo{volume}{24}},
  \bibinfo{pages}{045303} (\bibinfo{year}{2013}).

\bibitem[{\citenamefont{Suzuki et~al.}(2011)\citenamefont{Suzuki, Sasaki,
  Oikawa, Shiraishi, Suzuki, and Noguchi}}]{Suzuki:2011}
\bibinfo{author}{\bibfnamefont{T.}~\bibnamefont{Suzuki}},
  \bibinfo{author}{\bibfnamefont{T.}~\bibnamefont{Sasaki}},
  \bibinfo{author}{\bibfnamefont{T.}~\bibnamefont{Oikawa}},
  \bibinfo{author}{\bibfnamefont{M.}~\bibnamefont{Shiraishi}},
  \bibinfo{author}{\bibfnamefont{Y.}~\bibnamefont{Suzuki}}, \bibnamefont{and}
  \bibinfo{author}{\bibfnamefont{K.}~\bibnamefont{Noguchi}},
  \bibinfo{journal}{Appl. Phys. Expr.} \textbf{\bibinfo{volume}{4}},
  \bibinfo{pages}{023003} (\bibinfo{year}{2011}).

\bibitem[{\citenamefont{Wolfowicz et~al.}(2013)\citenamefont{Wolfowicz,
  Tyryshkin, George, Riemann, Abrosimov, Becker, Pohl, Thewalt, Lyon, and
  Morton}}]{Wolfowicz:2013}
\bibinfo{author}{\bibfnamefont{G.}~\bibnamefont{Wolfowicz}},
  \bibinfo{author}{\bibfnamefont{A.~M.} \bibnamefont{Tyryshkin}},
  \bibinfo{author}{\bibfnamefont{R.~E.} \bibnamefont{George}},
  \bibinfo{author}{\bibfnamefont{H.}~\bibnamefont{Riemann}},
  \bibinfo{author}{\bibfnamefont{N.~V.} \bibnamefont{Abrosimov}},
  \bibinfo{author}{\bibfnamefont{P.}~\bibnamefont{Becker}},
  \bibinfo{author}{\bibfnamefont{H.-J.} \bibnamefont{Pohl}},
  \bibinfo{author}{\bibfnamefont{M.~L.~W.} \bibnamefont{Thewalt}},
  \bibinfo{author}{\bibfnamefont{S.~A.} \bibnamefont{Lyon}}, \bibnamefont{and}
  \bibinfo{author}{\bibfnamefont{J.~J.~L.} \bibnamefont{Morton}},
  \bibinfo{journal}{Nat. Nanotechnol.} \textbf{\bibinfo{volume}{8}},
  \bibinfo{pages}{561} (\bibinfo{year}{2013}).

\bibitem[{\citenamefont{Goh et~al.}(2004)\citenamefont{Goh, Oberbeck, Simmons,
  Hamilton, and Clark}}]{Goh:2004}
\bibinfo{author}{\bibfnamefont{K.~E.~J.} \bibnamefont{Goh}},
  \bibinfo{author}{\bibfnamefont{L.}~\bibnamefont{Oberbeck}},
  \bibinfo{author}{\bibfnamefont{M.~Y.} \bibnamefont{Simmons}},
  \bibinfo{author}{\bibfnamefont{A.~R.} \bibnamefont{Hamilton}},
  \bibnamefont{and} \bibinfo{author}{\bibfnamefont{R.~G.} \bibnamefont{Clark}},
  \bibinfo{journal}{Appl. Phys. Lett.} \textbf{\bibinfo{volume}{85}},
  \bibinfo{pages}{4953} (\bibinfo{year}{2004}).

\bibitem[{\citenamefont{Goh et~al.}(2006)\citenamefont{Goh, Oberbeck, Simmons,
  Hamilton, and Butcher}}]{Goh:2006}
\bibinfo{author}{\bibfnamefont{K.~E.~J.} \bibnamefont{Goh}},
  \bibinfo{author}{\bibfnamefont{L.}~\bibnamefont{Oberbeck}},
  \bibinfo{author}{\bibfnamefont{M.~Y.} \bibnamefont{Simmons}},
  \bibinfo{author}{\bibfnamefont{A.~R.} \bibnamefont{Hamilton}},
  \bibnamefont{and} \bibinfo{author}{\bibfnamefont{M.~J.}
  \bibnamefont{Butcher}}, \bibinfo{journal}{Phys. Rev. B}
  \textbf{\bibinfo{volume}{73}}, \bibinfo{pages}{035401}
  (\bibinfo{year}{2006}).

\bibitem[{\citenamefont{Goh and Simmons}(2009)}]{Goh:2009}
\bibinfo{author}{\bibfnamefont{K.~E.~J.} \bibnamefont{Goh}} \bibnamefont{and}
  \bibinfo{author}{\bibfnamefont{M.~Y.} \bibnamefont{Simmons}},
  \bibinfo{journal}{Appl. Phys. Lett.} \textbf{\bibinfo{volume}{95}},
  \bibinfo{pages}{142104} (\bibinfo{year}{2009}). 

\bibitem[{\citenamefont{Carter et~al.}(2009)\citenamefont{Carter, Warschkow,
  Marks, and McKenzie}}]{Carter:2009a}
\bibinfo{author}{\bibfnamefont{D.~J.} \bibnamefont{Carter}},
  \bibinfo{author}{\bibfnamefont{O.}~\bibnamefont{Warschkow}},
  \bibinfo{author}{\bibfnamefont{N.~A.} \bibnamefont{Marks}}, \bibnamefont{and}
  \bibinfo{author}{\bibfnamefont{D.~R.} \bibnamefont{McKenzie}},
  \bibinfo{journal}{Phys. Rev. B} \textbf{\bibinfo{volume}{79}},
  \bibinfo{pages}{033204} (\bibinfo{year}{2009}).

\bibitem[{\citenamefont{Carter et~al.}(2011)\citenamefont{Carter, Marks,
  Warschkow, and McKenzie}}]{Carter:2011}
\bibinfo{author}{\bibfnamefont{D.~J.} \bibnamefont{Carter}},
  \bibinfo{author}{\bibfnamefont{N.~A.} \bibnamefont{Marks}},
  \bibinfo{author}{\bibfnamefont{O.}~\bibnamefont{Warschkow}},
  \bibnamefont{and} \bibinfo{author}{\bibfnamefont{D.~R.}
  \bibnamefont{McKenzie}}, \bibinfo{journal}{Nanotechnology}
  \textbf{\bibinfo{volume}{22}}, \bibinfo{pages}{065701}
  (\bibinfo{year}{2011}).

\bibitem[{\citenamefont{Miwa et~al.}(2013)\citenamefont{Miwa, Hofmann, Simmons,
  and Wells}}]{Miwa:2013}
\bibinfo{author}{\bibfnamefont{J.~A.} \bibnamefont{Miwa}},
  \bibinfo{author}{\bibfnamefont{P.}~\bibnamefont{Hofmann}},
  \bibinfo{author}{\bibfnamefont{M.~Y.} \bibnamefont{Simmons}},
  \bibnamefont{and} \bibinfo{author}{\bibfnamefont{J.~W.} \bibnamefont{Wells}},
  \bibinfo{journal}{Phys. Rev. Lett.} \textbf{\bibinfo{volume}{110}},
  \bibinfo{pages}{136801} (\bibinfo{year}{2013}).

\bibitem[{\citenamefont{Polley et~al.}(2012)\citenamefont{Polley, Clarke, Miwa,
  Simmons, and Wells}}]{Polley:2012a}
\bibinfo{author}{\bibfnamefont{C.~M.} \bibnamefont{Polley}},
  \bibinfo{author}{\bibfnamefont{W.~R.} \bibnamefont{Clarke}},
  \bibinfo{author}{\bibfnamefont{J.~A.} \bibnamefont{Miwa}},
  \bibinfo{author}{\bibfnamefont{M.~Y.} \bibnamefont{Simmons}},
  \bibnamefont{and} \bibinfo{author}{\bibfnamefont{J.~W.} \bibnamefont{Wells}},
  \bibinfo{journal}{Appl. Phys. Lett.} \textbf{\bibinfo{volume}{101}},
  \bibinfo{eid}{262105} (\bibinfo{year}{2012}).

\bibitem[{\citenamefont{Polley et~al.}(2013)\citenamefont{Polley, Clarke, Miwa,
  Scappucci, Wells, Jaeger, Bischof, Reidy, Gorman, and
  Simmons}}]{Polley:2013a}
\bibinfo{author}{\bibfnamefont{C.~M.} \bibnamefont{Polley}},
  \bibinfo{author}{\bibfnamefont{W.~R.} \bibnamefont{Clarke}},
  \bibinfo{author}{\bibfnamefont{J.~A.} \bibnamefont{Miwa}},
  \bibinfo{author}{\bibfnamefont{G.}~\bibnamefont{Scappucci}},
  \bibinfo{author}{\bibfnamefont{J.~W.} \bibnamefont{Wells}},
  \bibinfo{author}{\bibfnamefont{D.~L.} \bibnamefont{Jaeger}},
  \bibinfo{author}{\bibfnamefont{M.~R.} \bibnamefont{Bischof}},
  \bibinfo{author}{\bibfnamefont{R.~F.} \bibnamefont{Reidy}},
  \bibinfo{author}{\bibfnamefont{B.~P.} \bibnamefont{Gorman}},
  \bibnamefont{and} \bibinfo{author}{\bibfnamefont{M.}~\bibnamefont{Simmons}},
  \bibinfo{journal}{ACS Nano} \textbf{\bibinfo{volume}{7}},
  \bibinfo{pages}{5499} (\bibinfo{year}{2013}).

\bibitem[{\citenamefont{Miwa et~al.}(2014)\citenamefont{Miwa, Warschkow,
  Carter, Marks, Mazzola, Simmons, and Wells}}]{Miwa:2014a}
\bibinfo{author}{\bibfnamefont{J.~A.} \bibnamefont{Miwa}},
  \bibinfo{author}{\bibfnamefont{O.}~\bibnamefont{Warschkow}},
  \bibinfo{author}{\bibfnamefont{D.~J.} \bibnamefont{Carter}},
  \bibinfo{author}{\bibfnamefont{N.~A.} \bibnamefont{Marks}},
  \bibinfo{author}{\bibfnamefont{F.}~\bibnamefont{Mazzola}},
  \bibinfo{author}{\bibfnamefont{M.~Y.} \bibnamefont{Simmons}},
  \bibnamefont{and} \bibinfo{author}{\bibfnamefont{J.~W.} \bibnamefont{Wells}},
  \bibinfo{journal}{Nano Lett.} \textbf{\bibinfo{volume}{14}},
  \bibinfo{pages}{1515} (\bibinfo{year}{2014}).

\bibitem[{\citenamefont{Mazzola
  et~al.}(2014{\natexlab{a}})\citenamefont{Mazzola, Polley, Miwa, Simmons, and
  Wells}}]{Mazzola:2014}
\bibinfo{author}{\bibfnamefont{F.}~\bibnamefont{Mazzola}},
  \bibinfo{author}{\bibfnamefont{C.~M.} \bibnamefont{Polley}},
  \bibinfo{author}{\bibfnamefont{J.~A.} \bibnamefont{Miwa}},
  \bibinfo{author}{\bibfnamefont{M.~Y.} \bibnamefont{Simmons}},
  \bibnamefont{and} \bibinfo{author}{\bibfnamefont{J.~W.} \bibnamefont{Wells}},
  \bibinfo{journal}{Appl. Phys. Lett.} \textbf{\bibinfo{volume}{104}},
  \bibinfo{eid}{173108} (\bibinfo{year}{2014}{\natexlab{a}}).

\bibitem[{\citenamefont{Mazzola
  et~al.}(2014{\natexlab{b}})\citenamefont{Mazzola, Edmonds, H{\o}ydalsvik,
  Carter, Marks, Cowie, Thomsen, Miwa, Simmons, and Wells}}]{Mazzola:2014a}
\bibinfo{author}{\bibfnamefont{F.}~\bibnamefont{Mazzola}},
  \bibinfo{author}{\bibfnamefont{M.~T.} \bibnamefont{Edmonds}},
  \bibinfo{author}{\bibfnamefont{K.}~\bibnamefont{H{\o}ydalsvik}},
  \bibinfo{author}{\bibfnamefont{D.~J.} \bibnamefont{Carter}},
  \bibinfo{author}{\bibfnamefont{N.~A.} \bibnamefont{Marks}},
  \bibinfo{author}{\bibfnamefont{B.~C.~C.} \bibnamefont{Cowie}},
  \bibinfo{author}{\bibfnamefont{L.}~\bibnamefont{Thomsen}},
  \bibinfo{author}{\bibfnamefont{J.}~\bibnamefont{Miwa}},
  \bibinfo{author}{\bibfnamefont{M.~Y.} \bibnamefont{Simmons}},
  \bibnamefont{and} \bibinfo{author}{\bibfnamefont{J.~W.} \bibnamefont{Wells}},
  \bibinfo{journal}{ACS Nano} \textbf{\bibinfo{volume}{8}},
  \bibinfo{pages}{10223} (\bibinfo{year}{2014}{\natexlab{b}}).

\bibitem[{\citenamefont{Mazzola et~al.}()\citenamefont{Mazzola, Miwa, Rahman,
  Zhu, Simmons, Hofmann, and Wells}}]{Mazzola:2016}
\bibinfo{author}{\bibfnamefont{F.}~\bibnamefont{Mazzola}},
  \bibinfo{author}{\bibfnamefont{J.~A.} \bibnamefont{Miwa}},
  \bibinfo{author}{\bibfnamefont{R.}~\bibnamefont{Rahman}},
  \bibinfo{author}{\bibfnamefont{X.-G.} \bibnamefont{Zhu}},
  \bibinfo{author}{\bibfnamefont{M.}~\bibnamefont{Simmons}},
  \bibinfo{author}{\bibfnamefont{P.}~\bibnamefont{Hofmann}}, \bibnamefont{and}
  \bibinfo{author}{\bibfnamefont{J.}~\bibnamefont{Wells}}, \bibinfo{journal}{in
  progress}.

\bibitem[{\citenamefont{Jensen et~al.}(1997)\citenamefont{Jensen, Butorin,
  Kaurila, Nyholm, and Johansson}}]{Jensen:1997}
\bibinfo{author}{\bibfnamefont{B.}~\bibnamefont{Jensen}},
  \bibinfo{author}{\bibfnamefont{S.}~\bibnamefont{Butorin}},
  \bibinfo{author}{\bibfnamefont{T.}~\bibnamefont{Kaurila}},
  \bibinfo{author}{\bibfnamefont{R.}~\bibnamefont{Nyholm}}, \bibnamefont{and}
  \bibinfo{author}{\bibfnamefont{L.}~\bibnamefont{Johansson}},
  \bibinfo{journal}{Nucl. Instrum. Methods in Phys. Res. A}
  \textbf{\bibinfo{volume}{394}}, \bibinfo{pages}{243 } (\bibinfo{year}{1997}).

\bibitem[{\citenamefont{Curson et~al.}(2004)\citenamefont{Curson, Schofield,
  Simmons, Oberbeck, O'Brien, and Clark}}]{Curson:2004}
\bibinfo{author}{\bibfnamefont{N.~J.} \bibnamefont{Curson}},
  \bibinfo{author}{\bibfnamefont{S.~R.} \bibnamefont{Schofield}},
  \bibinfo{author}{\bibfnamefont{M.~Y.} \bibnamefont{Simmons}},
  \bibinfo{author}{\bibfnamefont{L.}~\bibnamefont{Oberbeck}},
  \bibinfo{author}{\bibfnamefont{J.~L.} \bibnamefont{O'Brien}},
  \bibnamefont{and} \bibinfo{author}{\bibfnamefont{R.~G.} \bibnamefont{Clark}},
  \bibinfo{journal}{Phys. Rev. B} \textbf{\bibinfo{volume}{69}},
  \bibinfo{pages}{195303} (\bibinfo{year}{2004}).

\bibitem[{\citenamefont{Wilson et~al.}(2006)\citenamefont{Wilson, Warschkow,
  Marks, Curson, Schofield, Reusch, Radny, Smith, McKenzie, and
  Simmons}}]{Wilson:2006}
\bibinfo{author}{\bibfnamefont{H.~F.} \bibnamefont{Wilson}},
  \bibinfo{author}{\bibfnamefont{O.}~\bibnamefont{Warschkow}},
  \bibinfo{author}{\bibfnamefont{N.~A.} \bibnamefont{Marks}},
  \bibinfo{author}{\bibfnamefont{N.~J.} \bibnamefont{Curson}},
  \bibinfo{author}{\bibfnamefont{S.~R.} \bibnamefont{Schofield}},
  \bibinfo{author}{\bibfnamefont{T.~C.~G.} \bibnamefont{Reusch}},
  \bibinfo{author}{\bibfnamefont{M.~W.} \bibnamefont{Radny}},
  \bibinfo{author}{\bibfnamefont{P.~V.} \bibnamefont{Smith}},
  \bibinfo{author}{\bibfnamefont{D.~R.} \bibnamefont{McKenzie}},
  \bibnamefont{and} \bibinfo{author}{\bibfnamefont{M.~Y.}
  \bibnamefont{Simmons}}, \bibinfo{journal}{Phys. Rev. B}
  \textbf{\bibinfo{volume}{74}}, \bibinfo{pages}{195310}
  (\bibinfo{year}{2006}).

\bibitem[{\citenamefont{Johansson
  et~al.}(1990{\natexlab{a}})\citenamefont{Johansson, Persson, Karlsson, and
  Uhrberg}}]{Johansson:1990a}
\bibinfo{author}{\bibfnamefont{L.~S.~O.} \bibnamefont{Johansson}},
  \bibinfo{author}{\bibfnamefont{P.~E.~S.} \bibnamefont{Persson}},
  \bibinfo{author}{\bibfnamefont{U.~O.} \bibnamefont{Karlsson}},
  \bibnamefont{and} \bibinfo{author}{\bibfnamefont{R.~I.~G.}
  \bibnamefont{Uhrberg}}, \bibinfo{journal}{Phys. Rev. B}
  \textbf{\bibinfo{volume}{42}}, \bibinfo{pages}{8991}
  (\bibinfo{year}{1990}{\natexlab{a}}).

\bibitem[{\citenamefont{Johansson
  et~al.}(1990{\natexlab{b}})\citenamefont{Johansson, Uhrberg, M\aa{}rtensson,
  and Hansson}}]{Johansson:1990}
\bibinfo{author}{\bibfnamefont{L.~S.~O.} \bibnamefont{Johansson}},
  \bibinfo{author}{\bibfnamefont{R.~I.~G.} \bibnamefont{Uhrberg}},
  \bibinfo{author}{\bibfnamefont{P.}~\bibnamefont{M\aa{}rtensson}},
  \bibnamefont{and} \bibinfo{author}{\bibfnamefont{G.~V.}
  \bibnamefont{Hansson}}, \bibinfo{journal}{Phys. Rev. B}
  \textbf{\bibinfo{volume}{42}}, \bibinfo{pages}{1305}
  (\bibinfo{year}{1990}{\natexlab{b}}).

\bibitem[{\citenamefont{Zhang et~al.}(2011)\citenamefont{Zhang, Richard, Qian,
  Xu, Dai, and Ding}}]{Zhang:2011}
\bibinfo{author}{\bibfnamefont{P.}~\bibnamefont{Zhang}},
  \bibinfo{author}{\bibfnamefont{P.}~\bibnamefont{Richard}},
  \bibinfo{author}{\bibfnamefont{T.}~\bibnamefont{Qian}},
  \bibinfo{author}{\bibfnamefont{Y.-M.} \bibnamefont{Xu}},
  \bibinfo{author}{\bibfnamefont{X.}~\bibnamefont{Dai}}, \bibnamefont{and}
  \bibinfo{author}{\bibfnamefont{H.}~\bibnamefont{Ding}},
  \bibinfo{journal}{Rev. of Sci. Instrum.}
  \textbf{\bibinfo{volume}{82}}, \bibinfo{pages}{043712}
  (\bibinfo{year}{2011}).

\bibitem[{SM1()}]{SM1}
\bibinfo{title}{See Supplementary Material for details regarding the fit
  analysis.}

\bibitem[{\citenamefont{Yengui et~al.}(2015)\citenamefont{Yengui, Pinto,
  Leszczynski, and Riedel}}]{Yengui:2015aa}
\bibinfo{author}{\bibfnamefont{M.}~\bibnamefont{Yengui}},
  \bibinfo{author}{\bibfnamefont{H.~P.} \bibnamefont{Pinto}},
  \bibinfo{author}{\bibfnamefont{J.}~\bibnamefont{Leszczynski}},
  \bibnamefont{and} \bibinfo{author}{\bibfnamefont{D.}~\bibnamefont{Riedel}},
  \bibinfo{journal}{J. of Phys.: Condens. Matter}
  \textbf{\bibinfo{volume}{27}}, \bibinfo{pages}{045001}
  (\bibinfo{year}{2015}).

\bibitem[{SM2()}]{SM2}
\bibinfo{title}{See Supplementary Material for details regarding the
  estimates of the effective masses.}

\bibitem[{\citenamefont{Bianchi et~al.}(2011)\citenamefont{Bianchi, Hatch, Mi,
  Iversen, and Hofmann}}]{Bianchi:2011}
\bibinfo{author}{\bibfnamefont{M.}~\bibnamefont{Bianchi}},
  \bibinfo{author}{\bibfnamefont{R.~C.} \bibnamefont{Hatch}},
  \bibinfo{author}{\bibfnamefont{J.}~\bibnamefont{Mi}},
  \bibinfo{author}{\bibfnamefont{B.~B.} \bibnamefont{Iversen}},
  \bibnamefont{and} \bibinfo{author}{\bibfnamefont{P.}~\bibnamefont{Hofmann}},
  \bibinfo{journal}{Phys. Rev. Lett.} \textbf{\bibinfo{volume}{107}},
  \bibinfo{pages}{086802} (\bibinfo{year}{2011}).

\end{thebibliography}



\end{document}


\title{Simultaneous conduction and valence band quantisation in ultra-shallow, high density doping profiles in semiconductors}

\author{F.~Mazzola}
\affiliation{Center for Quantum Spintronics, Department of Physics, Norwegian University of Science and Technology, NO-7491 Trondheim, Norway }
\author{J.~W.~Wells}
\affiliation{Center for Quantum Spintronics, Department of Physics, Norwegian University of Science and Technology, NO-7491 Trondheim, Norway }
\author{A.~C.~Pakpour-Tabrizi}
\affiliation{London Centre for Nanotechnology and Department of Electronic and Electrical Engineering, University College London, 17-19 Gordon Street, London WC1H 0AH, U.K.}
\author{R.~B.~Jackman}
\affiliation{London Centre for Nanotechnology and Department of Electronic and Electrical Engineering, University College London, 17-19 Gordon Street, London WC1H 0AH, U.K.}
\author{B.~Thiagarajan}
\affiliation{MAX-lab, PO Box 118, S-22100 Lund, Sweden}
\author{Ph.~Hofmann}
\affiliation{ Department of Physics and Astronomy and Interdisciplinary Nanoscience Center (iNANO), Aarhus University Ny Munkegade 120 DK-8000 Aarhus}
\author{J.~A.~Miwa}
\affiliation{ Department of Physics and Astronomy and Interdisciplinary Nanoscience Center (iNANO), Aarhus University Ny Munkegade 120 DK-8000 Aarhus}

\date{\today}


 \maketitle

\noindent \textbf{Details regarding the fit analysis of the quantised valence band states, qw$_1$ and qw$_2$, for different Si encapsulation thicknesses}
 \\

In this section we provide details on how a combination of energy distribution curves (EDCs) and momentum distribution curves (MDCs) were used to determine the positions of the quantised valence band states, qw$_1$ and qw$_2$.  Fig. \ref{fig:1} shows the angle resolved photoemission spectroscopy (ARPES) data for three $\delta$-layer samples, each with a different silicon encapsulation thickness: 2\,nm, 3\,nm and 4\,nm. The positions of the bands were determined by extracting EDCs and MDCs. The EDCs and MDCs were fitted with a Lorenztian convoluted with a Gaussian in order to extract the peak positions, which were then marked by green crosses and overlaid on the ARPES spectra shown in Fig. \ref{fig:1}.  The green crosses were then fitted with an even sixth-order polynomial,

\begin{equation}
E = a+bx^2+cx^4+dx^6
\end{equation}\\
\noindent in order to capture the quantised valence band positions. In this equation,  $a$ is the valence band maximum, and $b$, $c$ and $d$ are coefficients to fit the band dispersions. The resulting fits to the peak positions, extracted from the EDCs and MDCs, for the qw$_1$ (orange) and qw$_2$ (yellow) states are shown here in Fig. \ref{fig:1} and in Fig. 2 of the main paper. With regard to the 3\,nm and 4\,nm cases, the top of the qw$_1$ band, near k$_\parallel =$ 0,  is not visible because of a nearby surface state residing at a binding energy of $\approx$\,1\,eV, while for the qw$_2$ state the top of the band is readily visible.  On the other hand, the bands are sharper for the qw$_1$ state compared to those of the qw$_2$ state for an approximate  range of 0.2 $<$k$_\parallel$$<$ 0.5. The differences in  broadness and intensity of the bands account for the different number and positions of extracted EDCs and MDCs seen in Fig. \ref{fig:1}.

\begin{figure*}
\centering
\includegraphics[width=16cm]{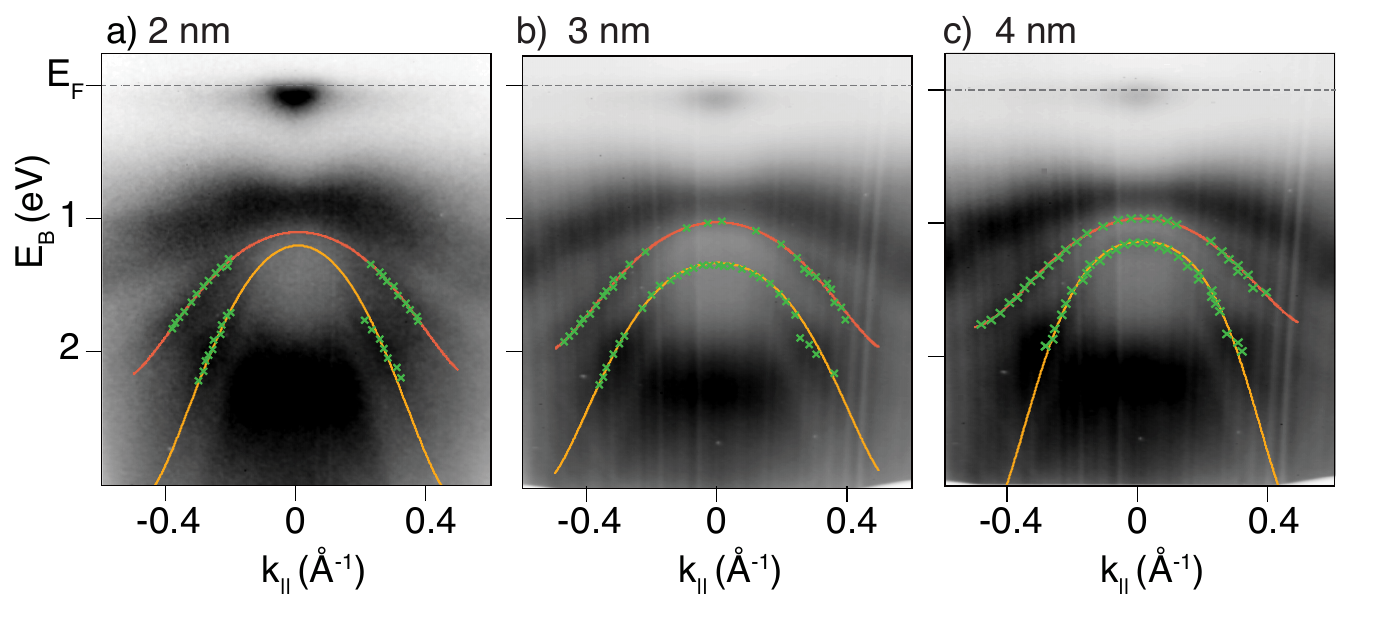}
\caption{ARPES data for three $\delta$-layer samples, each with a different silicon encapsulation thicknesses: (a) 2\,nm, (b) 3\,nm and (c) 4\,nm. The green crosses represent the peak maxima of the quantised valence bands extracted from EDCs and MDCs.  The peak positions were then fitted with an even sixth-order polynomial: qw$_1$ (orange) and qw$_2$ (yellow).} 
  \label{fig:1}
\end{figure*}

\newpage
\maketitle
\noindent  \textbf{Determining the effective masses for the quantised valence band states, qw$_1$ and qw$_2$, for different silicon encapsulation thicknesses}
\\

Based on the fits presented in Fig. \ref{fig:1}, we determined effective mass values for qw$_1$ and qw$_2$ states near the band maxima, for three $\delta$-layer samples with different silicon encapsulation thicknesses: 2\,nm, 3\,nm and 4\,nm. From the curvature of the qw$_1$ state at k$_\parallel =$ 0,  we find effective masses of 0.82\,$m_e$\,$\pm$\,0.13, 0.91\,$m_e$\,$\pm$\,0.08 and 0.92\,$m_e$\,$\pm$\,0.08 corresponding to  silicon encapsulation thicknesses of 2\,nm, 3\,nm and 4\,nm.  Using the same method for the qw$_2$ state, we find effective masses of  0.34\,$m_e$\,$\pm$\,0.12, 0.70\,$m_e$\,$\pm$\,0.09 and 0.64\,$m_e$\,$\pm$\,0.08 for respective silicon encapsulation thicknesses of 2\,nm, 3\,nm and 4\,nm.  We note that the uncertainties, stated along with the determined effective masses, were propagated through the even sixth-order polynomial used to fit the quantised valence band states (see previous section for details). The effective masses and uncertainties for the different samples are summarised in Table \ \ref{TabS1}.  
\begin{table}[h]
\centering
\caption{Summary of effective masses and uncertainties (at the band maxima) determined for the quantised valence band states for different silicon encapsulation thicknesses.\\ }
\label{TabS1}
\begin{tabular}{|   c   |   c   |   c   |}
\hline
encapsulation (nm) & qw$_1$ & qw$_2$  \\
\hline
2 &   0.82\,$m_e$\,$\pm$\,0.13   &  0.34\,$m_e$\,$\pm$\,0.12   \\
3 & 0.91\,$m_e$\,$\pm$\,0.08 & 0.70\,$m_e$\,$\pm$\,0.09  \\
4 & 0.92\,$m_e$\,$\pm$\,0.08 & 0.64\,$m_e$\,$\pm$\,0.08  \\
\hline
\end{tabular}
\end{table}

Note that for the 2\,nm case the uncertainties associated with the effective mass values for qw$_1$ and qw$_2$ are larger than the uncertainties for the thicker Si encapsulations.  And, the effective mass for qw$_2$ for the 2\,nm case differs significantly from  the effective mass values for qw$_2$  determined for both the  3\,nm and 4\,nm cases.  The larger uncertainties associated with the 2\,nm encapsulation thickness is hardly surprising given that  the band positions could not be accurately determined for quantised valence band states in the region of -0.2 $<$k$_\parallel$$<$ 0.2; see Fig. \ref{fig:1}(a). Additionally, we provide a summary of the estimated positions of the band maxima for the quantised valence band states for the 2\,nm, 3\,nm and 4\,nm silicon encapsulation thicknesses in Table \ \ref{TabS2}. For the thinner Si encapsulation thicknesses (i.e 1\,nm and 2\,nm) it is likely that there is a more pronounced spatial overlap of the surface state at 1\,eV binding energy with the quantised valence band states, compared with the thicker encapsulations (i.e. 3\,nm and 4\,nm), that can manifest as broadening and energy shifts of the states. Such an effect can lead to a less accurate determination of the effective mass. 
\begin{table}[h]
\centering
\caption{Band maxima for the quantised valence bands states for different silicon encapsulation thicknesses. The last column of the table gives the separation energies between qw$_1$ and qw$_2$ for the different samples. Uncertainties are stated, except for the 2\,nm case. Due to the lack of green crosses near the top of the band maxima for the 2\,nm sample, see Fig. \ref{fig:1}(a), it was not possible to provide a reasonable uncertainty. \\ }
\label{TabS2}
\begin{tabular}{|   c   |   c   |   c   |   c   |}
\hline
encapsulation (nm) & qw$_1$ & qw$_2$  &  qw$_1 -$ qw$_2$ \\
\hline
2 &   1.1\,eV   & 1.2\,eV & 0.1\,eV  \\
3 & 1.02 $\pm$ 0.08\,eV & 1.32 $\pm$ 0.09\,eV  & 0.30 $\pm$ 0.17\,eV \\
4 & 1.02 $\pm$ 0.08\,eV & 1.19 $\pm$0.04\,eV  & 0.17 $\pm$ 0.12\,eV \\
\hline
\end{tabular}
\end{table}
 
We expect that the different silicon encapsulation thicknesses will have a small affect on the effective mass values for the quantised valence bands states, but given the magnitude of the associated uncertainties, the effective mass values for the qw$_1$ state are in agreement with the heavy-hole state in the bulk valence band of Si.  The effective mass values for the qw$_2$ state are somewhat less, but also probably derived from the bulk heavy-holes since their effective mass values are more similar to that of the bulk heavy-hole state than the bulk light-hole state.